Katarzyna PASIERB*

Tomasz KAJDANOWICZ**

Przemysław KAZIENKO***

# PRIVACY-PRESERVING DATA MINING, SHARING AND PUBLISHING

The goal of the paper is to present different approaches to privacy-preserving data sharing and publishing in the context of e-health care systems. In particular, the literature review on technical issues in privacy assurance and current real-life high complexity implementation of medical system that assumes proper data sharing mechanisms are presented in the paper. [I]

## 1. INTRODUCTION

In the constantly developing world, e-health systems hold great promise for improving global access to healthcare services. Current significant technological visions of innovation in healthcare systems identify an approach to join different technological sectors and the need for technological platforms as well. These are: standardized electronic health records (eHR), aggregated public health data, genomic medicine, remote healthcare and diagnostics (telemedicine).

Advancements enable medical consultation, remote imaging services, specialized medical diagnostics, and etc. There is an increasing demand for good health data management. According to [16] 75% of Americans would like to communicate via e-mail with their physicians and 60% would like to track their medical records electronically. A nationwide system of electronic medical records promises to facilitate the exchange of medical knowledge and patient data among physicians and other health providers. The question is how can healthcare institutions share patient information with a third party without compromising the privacy of individual patients?

* Wroclaw University of Technology, katarzyna.pasierb@pwr.wroc.pl
** Wroclaw University of Technology, tomasz.kajdanowicz@pwr.wroc.pl
*** Wroclaw University of Technology, kazienko@pwr.wroc.pl

[I] This is not the final version of this paper. You can find the final version on the publisher web page.

At the beginning we recall terms privacy, confidentiality, and information security [25], [26]:
1. Privacy is the right of an individual to control disclosure of his or her medical information.
2. Confidentiality is the understanding that medical information will only be disclosed to authorised users at specific times of need. It entails holding sensitive data in a secure environment limited to an appropriate set of authorized individuals or organizations.
3. Information security includes the processes and mechanisms used to control the disclosure of information. It is the protection of computer-based information from unauthorized destruction, modification, or disclosure.

The privacy and security aspects have an effect on the electronic storage and transmitting of patient health information, see Figure 1. Vast quantities of data are generated through the health care process in medical institutions. We can distinguish different types of patient data: registration data (e.g., contact info), demographics (e.g., DOB, gender, race), billing information (e.g., diagnosis codes), genomic information (e.g., SNPs), medication and allergies, immunization status, laboratory test results, radiology images and so on. All kind of medical data connected with patient interacts in EMR System which consolidates particular systems, such as Registration System data (date and time of visit), Lab System, Pharmacy System, Radiology System (reports, images), Billing System (diagnosis codes), Order Entry System (prescriptions, orders), Decision Support System (clinical knowledge, guidelines). Physicians are the point of the transition/movement/usage of data. When they have access to all types of medical data related to the patients, can better diagnose and treat diseases with the help of Decision Support System. Medical information systems involve subsystems containing among others patient information, reporting tools, decision support systems and clinical scheduling.

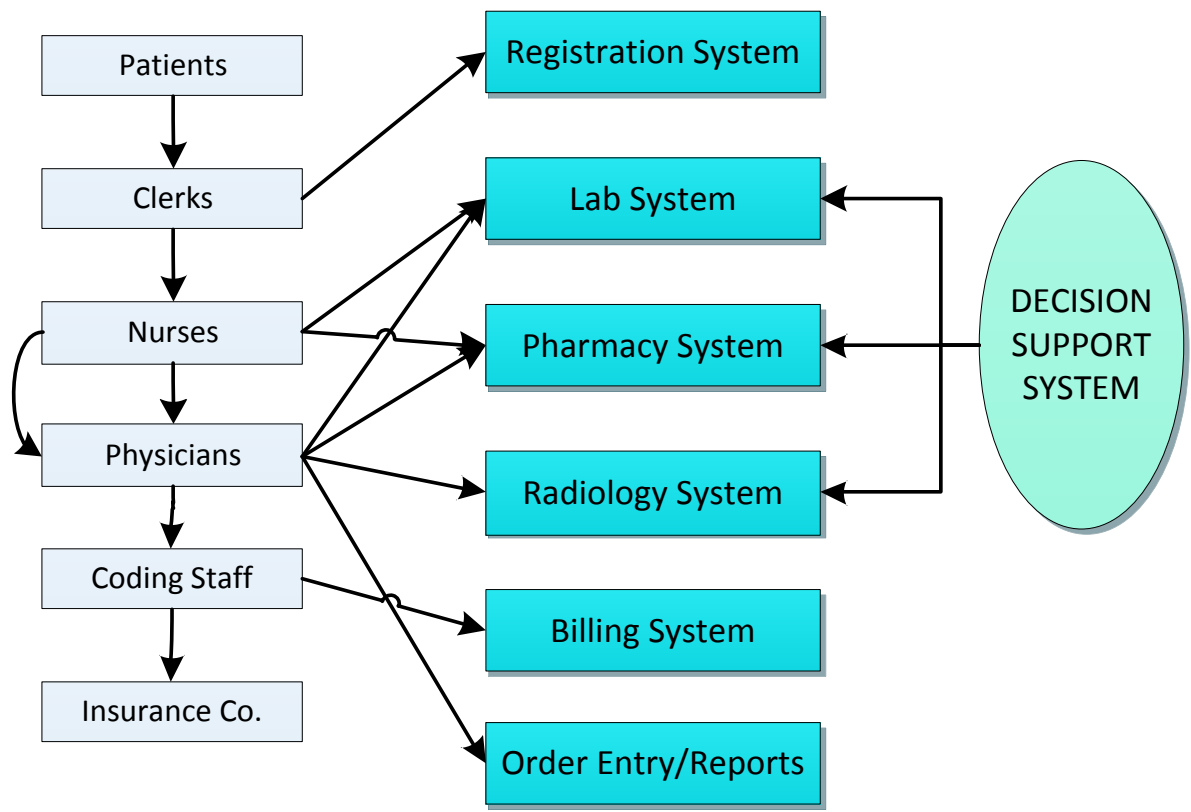

Figure 1 Different types of patient data

Such information exchange creates necessity of interoperability among healthcare systems. To fulfil those needs many of institutions introduced regulations to address specific concerns about privacy, security, and patient identification. During the last few years, the attention of governments around the world has been focused on transitioning the national health care system to an infrastructure based upon information technology [4]. A huge number of patient electronic records is available for mining, and in the meantime their privacy protection is be required by law what creates an demand for privacy-preserving data mining, sharing and publishing.

2. PRIVACY-PRESERVING DATA MINING (PPDM), PUBLISHING (PPDP),

Privacy-preserving data management is an important emerging research area that emerged in response to two important needs: data analysis and ensuring the privacy of the data owners. Privacy-preserving data publishing emphasizes the importance of need for privacy, threats in data sharing. A new approach seeks to protect data without focusing on the infrastructure level, but at element or aggregate data type. This type of pervasive security can be achieved by classifying data and enforcing access-control [2]. This process should be carried out in compliance with the constantly evolving regulations such as the European Union Data Protection Directive, the Health Insurance Portability and Accountability Act (HIPAA), California Senate Bill 1386, and industry standards such as the Payment Card Industry Data Security Standard. The Privacy Rule broadly defines „protected health information" as individually identifiable health information maintained or transmitted by a covered entity in any form or medium [1].

From a privacy viewpoint, the data attributes can be classified into three categories as below [13], [18], [19], [20]:
- Identity attributes, which can be used to directly identify an individual, including social security number, name, phone number, credit card number, address.
- Sensitive attributes, which contain private information that an individual typically does not want revealed: salary, medical test results, academic transcripts.
- Non-sensitive attributes, which are normally not considered as sensitive by individuals; many of these attributes can be found from publicly available sources Examples include age, gender, race, education, occupation, height, eye colour, and so on.

### 2.1. TYPICAL SCENARIO OF PRIVACY-PRESERVING DATA PUBLISHING

Privacy-preserving data publishing uses methods and tools both for publishing and preserving data privacy. In general there are three parties involved in the privacy problem in data mining:
- information owner - wants to discover knowledge from the data without compromising the confidentiality of the data;
- information providers (record owners) - individuals who provide their personal information to the data owner and want their privacy to be protected;
- data user/miner/recipient - has access to the data released by the data owner and can conduct data mining on the published. Data miner is considered as potential privacy intruder.

Basing on that, each scenario of data publishing has its own assumptions and requirements on the each of described parties. A typical scenario consists of two main phases:
1) data collection - the data publisher collects data from record owners;
2) data publishing - the data publisher releases the collected data to a data miner or the public;

In an untrusted model, a risk of revealing sensitive information from record exists. A data miner is not trusted and may try to identify sensitive information from the record owners. In that case we expect that a data publisher to do more than anonymizing the data.

In a basic form the data publisher has a table as follows: T(EI, QID, SA, N-SA)
- *Explicit Identifier* (EI) - a set of attributes: social security number containing information that explicitly identifies record owners;
- *Quasi Identifier* (QID) - a set of attributes that could potentially identify record owners;
- *Sensitive Attributes* (SA)- patient-specific information: disease, medical history, disability status;
- *Non - Sensitive Attributes* (N-SA) - all attributes that do not fall into the previous three categories

### 2.2. GENERAL CONCEPTS OF DATA MINING IN E-HEALTH

From the medical data utility point of view, data mining analysis can be performed over original data and over sanitized data. Sanitized and then shared or published data is referred to as privacy-preserving data publishing. Privacy-preserving data mining refers to the area of data mining that seeks to safeguard sensitive information from unsolicited or unsanctioned disclosure [4]. The term privacy-preserving data mining was introduced in 2000 in the papers [8] and [9] by mining a data set partitioned across several private enterprises.

### 2.3. ALGORITHMS AND METHODS

In this paragraph, an overview of the popular approaches for doing PPDS is presented. Anonymization algorithms enable transforming data in a way that satisfies privacy with minimal utility loss by using heuristics. Algorithms can be divided into partition-based and clustering-based. The first group treats a record as a multidimensional point with particular attributes. To split the data into datasets and highlight selected attribute (e. g. disease) Mondrian algorithm can be used. Other methods can be applied, such as R-tree based algorithm, optimized partitioning for intended tasks as classification, regression, query answering [5].

Privacy-preserving data solutions:
- **Synthetic data generation** - build a statistical model using a noise infused version of the data, and then synthetic data are generated by randomly sampling from this model
- **Masking methods** [13]
    - Perturbative - > randomization. The randomization approach [7] protects the patients' data by letting them randomly perturb their records before sending them to the server, taking away some real information and including some noise. Data miner's knowledge (belief) is modelled as a probability distribution. Main features:
        - The aim is to preserve privacy and aggregate statistics (e.g., means and correlation coefficients), falsify the data.
        - Methods: noise addition, data swapping, microaggregation, rounding.
        - One of the popular data-masking methods is noise-based perturbation [17]. The basic idea of this approach is to add noise to the sensitive data to disguise their true values, while preserving the statistical properties of the data.
        - Another popular data-masking approach is microaggregation which masks data by aggregating attribute values [14], [15].
      
      Pros:
        - privacy guarantees can be proven by just studying the randomization algorithm, not the data mining operations.
      
      Cons:
        - one of the limitations is that the data-masking methods apply to numeric data;
        - the results are always approximate; high-enough accuracy often requires a lot of randomized data.
    - Non-perturbative
        - aim at changing the granularity of the reported data;
        - do not falsify data.
- **Suppression -** withholding information due to disclosure constraints [4].
    - Record suppression - all values in a record are deleted prior to data publishing, results in excessive information loss,

- o Value suppression - certain values in quasi-identifiers are deleted (replaced by *) prior to data publishing or can be replaced with a less informative value by rounding (e.g. 55.22 to 50.00) or using intervals (e.g. 11-15, 16-20),
    - Cons: the analysis may be difficult if the choice of alternative suppressions depends on the data being suppressed, or if there is dependency between disclosed and suppressed data. Suppression cannot be used if data mining requires full access to the sensitive values.
- **Generalization** - the act of haphazardly perturbing data before disclosure [4], so when there are fewer distinct values data linkage becomes more difficult, e. g. address to zip code.
- **De-identification** - the process of altering the data set to limit identity linkage [4]. Data can be anonymized with different options including full de-identification, partial de-identification, and statistical anonymization based on $k$-anonymization. Full de-identification would render the data not very useful for many data analysis purposes. Partial de-identification provides better data utility. Statistical de-identification attempts to maintain as much ''useful'' data as possible while guaranteeing statistically acceptable data privacy. The concept of $k$-anonymity is described in Sec. 2.3.1.
- **Cryptography** - assumes that the data are stored at several private parties who agree to disclose the result of a certain data mining computation performed jointly over their data [4].
  The first adaptation of cryptographic techniques to data mining was done by [9] they introduced a decision tree construction over horizontally partitioned data; Cryptography challenge: scalability
- **Summarization** - releasing the data in the form of a summary that allows the evaluation of certain classes of aggregate queries while hiding the individual records Summarization extends randomization, but a summary is often expected to be much shorter [4].
  Many of data-masking methods due to high computational cost are directed at data presented in summarized tables, instead of a dataset of individuals usually required in data mining. That is not necessarily consistent with preserving data-mining quality.
  - o **Cons:** Verifying privacy guarantees for tabular data is challenging because of the potential for disclosure by inference [4].

### 2.3.1. *K*-ANONYMITY

One of the most popular methods recently has been $k$-anonymization. It aims at preventing sensitive information about individuals being identified or inferred from the dataset [5]. In case of $k$-anonymity, the system masks the values of some potentially identifying attributes, called quasi-identifier According to this principle each record in a relational table $T$ needs to have the same value over quasi-identifiers with at least $k$-1 other records in $T$. It is considered as a better protection than exposing all the information in the dataset.

Cons:
- $k$-anonymity is difficult to achieve before all data are collected in one trusted place [6];

- attributes that are not among quasi-identifiers, even if sensitive (e.g., diagnosis), are not suppressed and may get linked to an identity [6];

The query precision on the de-identified dataset using different de-identification options ($k$ parameter) is presented in Figure 2. The full de-identification provides the maximum privacy protection, but suffers a low query precision (57%). Statistical de-identification offers a guaranteed privacy level while maximizing the data utility. As expected, the larger $k$, the better the privacy level and the lower the query precision as the original data are generalized to a larger extent [21].

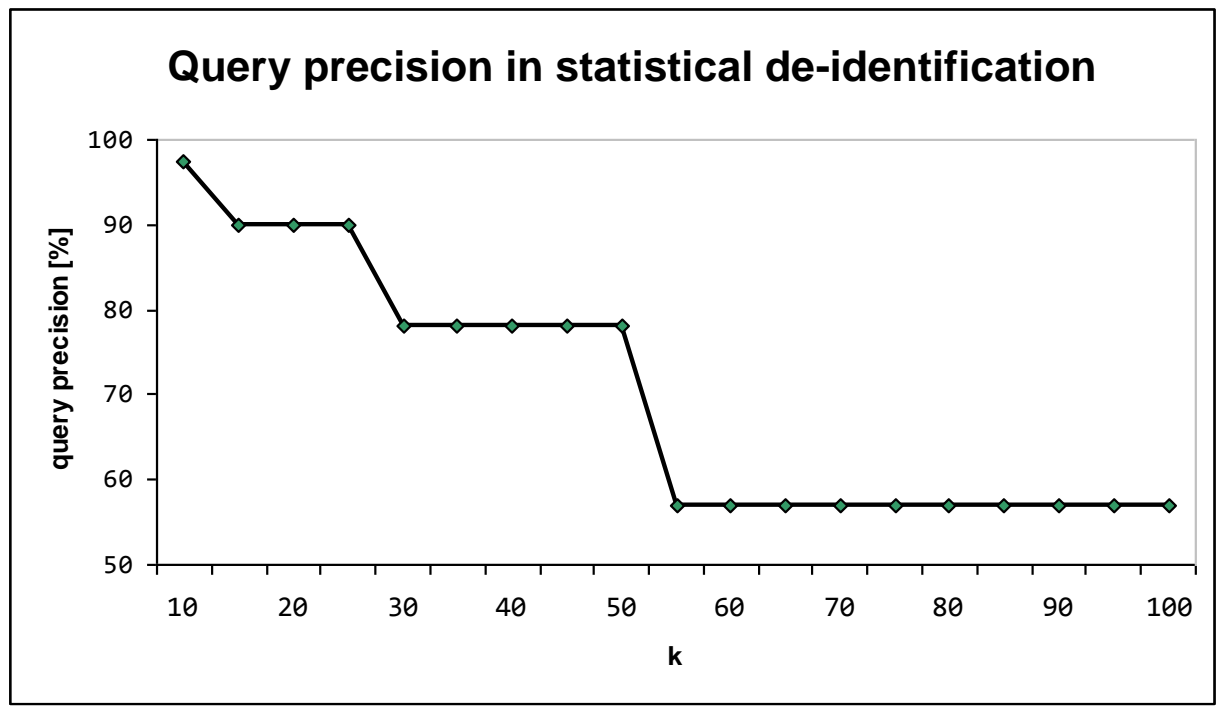

Figure 2 Query precision using statistical de-identification with respect to $k$; based on [21]

In a record linkage attack, the victim is vulnerable to being linked to the small number of records in the group. [22] presents a classification of record linkage anonymization algorithms and divides them into two groups: optimal anonymization and minimal anonymization. Algorithms in both groups use generalization and suppression to achieve the k-anonymity privacy model.

The first group finds an optimal $k$-anonymization by full-domain generalization and record suppression: MinGen algorithm with exhaustive search, binary search algorithm, optimal bottom-up generalization algorithms called Incognito, K-Optimize which uses the flexible subtree generalization is considered as an efficient optimal algorithm in this group.

The second family of algorithms employs a greedy search guided by a search metric. These, heuristic in nature, algorithms find a minimally anonymous solution, but are more scalable than the previous family: µ-argus algorithm, Top-Down Refinement or Iyengar's genetic algorithms application.

Another type of threat is connected with attribute linkage. Some sensitive values associated to the group can be easy to infer even if $k$-anonymity is satisfied. $l$-diversity, Confidence Bounding and $t$-closeness have been proposed to prevent attribute linkage.

## 3. PRIVACY-PRESERVING DATA SHARING IN POLISH ELECTRONIC PLATFORM FOR COLLECTION, ANALYSIS AND SHARING OF DIGITAL MEDICAL RECORDS

In years 2009 – 2014 National Centre of Health Information Systems (CSIOZ) has been the beneficiary of the P1 project – „Electronic Platform for Collection, Analysis and Sharing of Digital Medical Records (eHR, ePrescription, web portal, Patient's Internet Account)". The aim of the project is to provide the access to appropriate data (eHR, data warehouses, diagnoses, and other data repositories) and ensure interoperability and analyses among multiple medical systems.

At the current stage of P1 development the privacy preserving data sharing has been considered as a problem of access rights definition to information available to different categories of users. The authors of project's feasibility study [28], [29] suggest that an ideal solution would provide a possibility to associate each elementary item in the medical records of a person to people authorized to view this information, set up or approved by the person whom the data relates to. Such solution would naturally define dynamic groups of people responsible for care, provision of medical services or the protection of life for each patient. It could be extended to conditional statement of rights to access data from other reasons than the provision of medical care. This may relate to the management of health services, studies or research.

Project P1 identifies multiple difficulties associated with privacy preserving data sharing:
- a large number of items in medical records of patient care and high structural complexity of these data; thus difficult to classify the sensitivity of any item in medical records in a standardized manner,
- difficulty in determining how important is an individual position in medical records and may be relevant to each category of users,
- a large number of health care organizations gathering patients data
- the need of real-time appropriate access, in a distributed data processing environment to medical records,
- variety of security interest in groups of patients – there can be distinguished at least a group of people interested in high-level access control, relatively large, and a group of low level of interest in access control, representing most patients.

Summarizing, the overall medical system should ensure safe and consistent interchangeability of medical data between health care entities and is required to automate the negotiation whether a person contracting medical data should be authorized to receive it. In particular, the privacy preserving data sharing functionality should provide a safe, automated and rapid exchange of data through clear common understanding (agreed protocols, formats), safe and secure communication channels, straightforward identification of patients, health professionals and other interested in data acquisition entities. A basic solution for data sharing in P1 is based on access control list, including negotiated security policies of cooperating organizations and the definition of roles in the system.

## 4. CONCLUSIONS

The main technical challenge for PPDM is to make its algorithms scalable and achieve higher accuracy while keeping the privacy guarantees [4]. Another significant point is seamless integration within applications, databases and file formats. As it can be observed in currently developed medical information technology systems, architects and system designers should be more aware of the possibility of using methods and algorithms that solve the problem of privacy preserving data sharing.